\documentstyle[12pt]{article}

\begin{document}


\def\a{\alpha}
\def\b{\beta}
\def\c{\varepsilon}
\def\d{\delta}
\def\e{\epsilon}
\def\f{\phi}
\def\g{\gamma}
\def\h{\theta}
\def\k{\kappa}
\def\l{\lambda}
\def\m{\mu}
\def\n{\nu}
\def\p{\psi}
\def\q{\partial}
\def\r{\rho}
\def\s{\sigma}
\def\t{\tau}
\def\u{\upsilon}
\def\v{\varphi}
\def\w{\omega}
\def\x{\xi}
\def\y{\eta}
\def\z{\zeta}
\def\D{\Delta}
\def\G{\Gamma}
\def\H{\Theta}
\def\L{\Lambda}
\def\F{\Phi}
\def\P{\Psi}
\def\S{\Sigma}

\def\o{\over}
\def\beq{\begin{eqnarray}}
\def\eeq{\end{eqnarray}}
\newcommand{\gsim}{ \mathop{}_{\textstyle \sim}^{\textstyle >} }
\newcommand{\lsim}{ \mathop{}_{\textstyle \sim}^{\textstyle <} }

\def\IJMP{Int.~J.~Mod.~Phys. }
\def\MPL{Mod.~Phys.~Lett. }
\def\NP{Nucl.~Phys. }
\def\PL{Phys.~Lett. }
\def\PR{Phys.~Rev. }
\def\PRL{Phys.~Rev.~Lett. }
\def\PTP{Prog.~Theor.~Phys. }
\def\ZP{Z.~Phys. }


\baselineskip 0.7cm

\begin{titlepage}
\begin{flushright}
UT-905
\\
August, 2000
\end{flushright}

\vskip 1.35cm
\begin{center}
{\large \bf
Warped Compactification with an Abelian Gauge Theory
}
\vskip 1.2cm
S.~Hayakawa${}^1$ and Izawa K.-I.${}^{1,2}$
\vskip 0.4cm

${}^1${\it Department of Physics, University of Tokyo,\\
     Tokyo 113-0033, Japan}

${}^2${\it Research Center for the Early Universe, University of Tokyo,\\
     Tokyo 113-0033, Japan}

\vskip 1.5cm

\abstract{
We investigate warped compactification
with an abelian gauge theory in six dimensions.
The vanishing cosmological constant in four dimensions
can generically be realized with a regular metric
even in a 3-brane background without fine tuning of couplings.
}
\end{center}
\end{titlepage}

\setcounter{page}{2}

\section{Introduction}

The smallness of the cosmological constant poses
a severe problem
\cite{Wei}
on our natural understanding
of an effective field theory description of the universe.
The problem is two-fold:
one is the apparent absence (or cancellation) of the contributions
from the standard model dynamics including gravity
to the vacuum energy; the other is to choose the vanishing (or
tiny) value itself among possible values of the cosmological constant
even if we can choose it.

Warped compactification of a higher-dimensional theory
is an attempt to achieve the four-dimensional
vanishing cosmological constant without fine tuning of couplings
\cite{Rub}.
It results in a degenerate metric in six-dimensional pure gravity
\cite{Rub}.
Six-dimensional warped compactification
with an abelian gauge theory was investigated in
Ref.\cite{Wet}
and that with a 3-brane in
Ref.\cite{Cho}.

In this paper, we consider warped compactification
with an abelian gauge theory in a 3-brane background.
The vanishing four-dimensional cosmological constant 
can generically be realized
with regular and compact extra dimensions
in contrast to the case of pure gravity
without an abelian gauge theory.

\section{The Model}

Let us consider six-dimensional gravity coupled to an abelian gauge theory.
The action with a 3-brane is given by
\cite{Sun}
\beq
 S = \int \! d^6x \sqrt{-g}
     \left({1 \o 2}R - {1 \o 4}F_{MN}F^{MN} - \L \right)
   - \int \! d^4x \sqrt{-g_4} \l,
 \label{BA}
\eeq
where $g=\det g_{MN}$, $g_4=\det g_{\mu \nu}$, $\l > 0$,
the six-dimensional gravitational scale is set to unity,
and the brane is located at the origin in the extra two dimensions.
Here $M$ and $N$ denote six-dimensional indices,
$\mu$ and $\nu$ denote four-dimensional ones,
and $g_{\mu \nu}$ is the induced metric on the brane.
Then the Lagrangian in six dimensions is given by
\beq
 {\cal L} = \sqrt{-g}\left({1 \o 2}R - {1 \o 4}F_{MN}F^{MN} - \L \right)
          - \sqrt{-g_4} {\l \o 2\pi \e} \H(\e-r),
 \label{BL}
\eeq
where $\H$ denotes a step function and $\e=+0$.
Here we have adopted the polar coordinates $(r, \h)$
for the extra two dimensions ($0 \leq r$, $0 \leq \h < 2\pi$).

The equations of motion are obtained as
\beq
 & & R^{MN}-{1 \o 2}g^{MN}R = F^M{}_LF^{NL} - \left({1 \o 4}F_{KL}F^{KL}
     + \L \right)g^{MN} - \sqrt{g_4 \o g}
     \d^M_\mu \d ^N_\nu g^{\mu \nu}{\l \o 2\pi \e} \H(\e-r),
 \nonumber
 \\
 & & \q_M(\sqrt{-g}F^{MN}) = 0.
 \label{E0}
\eeq
In the following sections, we solve these equations
based on an ansatz of four-dimensionality after compactification.

\section{Warped Compactification}

In order to obtain four-dimensional spacetime from six dimensions,
we compactify the extra two dimensions.
Under an assumption of rotational symmetry
in the extra dimensions
($\h$-independence
\cite{Rub,Wet,Cho}
or orbifolding by $S^1$),
the warped metric is given by
\beq
 ds^2 = \s (r){\bar g}_{\mu \nu}dx^\mu dx^\nu - dr^2 - \r(r)d\h^2
 \label{WM}
\eeq
and the background gauge field is given by
\beq
 A_\mu = A_r = 0, \quad A_\h = a(r),
 \label{WG}
\eeq
where ${\bar g}_{\mu \nu}$ denotes the four-dimensional metric
independent of $(r, \h)$.

With the aid of Eq.(\ref{E0}),
we obtain
\beq
 F^{r \h} = {B \o \s^2 \sqrt{\r}},
\eeq
where $B$ is an integration constant taking continuous values.%
\footnote{The freedom to choose a value of one continuous variable $B$
eventually leads to that of
the effective four-dimensional cosmological constant $\L_4$.}
Then the Einstein equations are reduced to
\beq
 & & {3 \o 2}{\s'' \o \s}+{3 \o 4}{\s' \o \s}{\r' \o \r}
     -{1 \o 4}{\r'^2 \o \r^2}+{1 \o 2}{\r'' \o \r}
     = -{B^2 \o 2\s^4}
     -\L-{\l \o 2\pi \e \sqrt{\r}} \H(\e-r)+{\L_4 \o \s},
 \\
 & & {3 \o 2}{\s'^2 \o \s^2}+{\s' \o \s}{\r' \o \r}
     = {B^2 \o 2\s^4}-\L+{2\L_4 \o \s},
 \label{E1}
 \\
 & & 2{\s'' \o \s}+{1 \o 2}{\s'^2 \o \s^2}
     = {B^2 \o 2\s^4}-\L+{2\L_4 \o \s},
 \label{E2}
\eeq
where $\L_4$ denotes the four-dimensional cosmological constant
for the metric ${\bar g}_{\mu \nu}$
introduced as an integration constant
and the prime denotes a derivative with respect to $r$.

By means of Eq.(\ref{E1}) and (\ref{E2}), we obtain
\beq
 & & z'' = -{\q V \o \q z}; \quad
 V(z) = {25 \o 96}B^2 z^{-{6 \o 5}}+{5 \o 16}\L z^2
      - {25 \o 24}\L_4 z^{6 \o 5},
 \label{E3}
 \\
 & & \s = z^{4/5}, \quad \r = C^{-2}z'^2z^{-6/5},
 \label{E4}
\eeq
where $C$ is an integration constant.
Note that Eq.(\ref{E3})
looks like an equation of motion for a particle with the position
$z(r)$ at the time $r$ in a potential $V(z)$
\cite{Rub,Wet}.

\section{The Solutions}

Let us seek regular metrics with $\L_4=0$. From the
Einstein equations in the previous section,
boundary conditions at $r=\e$ are given by
\cite{Rub,Wet,Cho}
\beq
 (\sqrt{\r})'|_{r=0}^\e = -{\l \o 2\pi},
 \quad (\sqrt{\r})'(0)=1,\quad \r(\e)=0.
\eeq
Namely, the extra dimensions are conical around the brane
with a deficit angle $\l$, which is regular
in the presence of the brane.
Owing to Eq.(\ref{E4}), this implies
\beq
 z'(\e)=0, \quad z''(\e)=C\left(1-{\l \o 2\pi}\right).
\eeq
Here we have taken $z(\e)=1$ without loss of generality.
Then Eq.(\ref{E3}) leads to
\beq
 C\left(1-{\l \o 2\pi}\right)=z''(\e)=-{\q V \o \q z}(z(\e))
 ={5 \o 16}B^2-{5 \o 8}\L.
 \label{F1}
\eeq

For $\L > 0$ and $\l < 2\pi$, we have desired solutions
of Eq.(\ref{E3}), which are half periods of
oscillations in $z(r)$
between two values $1=z(\e)$ and ${\bar z}=z({\bar r})$
with $z'({\bar r})=0$ given by
\beq
 {25 \o 96}B^2 + {5 \o 16}\L=V(z(\e))=V(z({\bar r}))
 ={25 \o 96}B^2 {\bar z}^{-{6 \o 5}} + {5 \o 16}\L {\bar z}^2.
 \label{F2}
\eeq
Imposing a condition
$(\sqrt{\r})'({\bar r})=-1$ for a regular metric
(without a conical singularity)
at $r = {\bar r}$, we obtain
\beq
 -C{\bar z}^{3 \o 5}=z''({\bar r})=-{\q V \o \q z}(z({\bar r}))
 ={5 \o 16}B^2{\bar z}^{-{11 \o 5}}-{5 \o 8}\L{\bar z}.
 \label{F3}
\eeq

The constants $B$, $C$ and ${\bar z}$
are determined by Eq.(\ref{F1}), (\ref{F2}) and (\ref{F3}).
This result indicates that we indeed have compact extra dimensions
of sphere topology with a completely regular metric in six dimensions.

\section{Conclusion}

We have considered warped compactification Eq.(\ref{WM}) and (\ref{WG})
with an abelian gauge theory in a 3-brane background
Eq.(\ref{BA}) or (\ref{BL}).
The vanishing cosmological constant $\L_4=0$ can generically be realized
with a regular metric determined by
Eq.(\ref{F1}), (\ref{F2}) and (\ref{F3})
with Eq.(\ref{E3}) and (\ref{E4}).
This is achieved without fine tuning of Lagrangian parameters.

We have only obtained the backgrounds with $\L_4=0$,
just for simplicity;
the backgrounds with $\L_4 \neq 0$ are also possible.
The four-dimensional cosmological constant $\L_4$
is an integration constant in the present setup.
This diminishes a half of the cosmological constant problem
stated in the Introduction:
the desired backgrounds with $\L_4=0$ exist
for positive couplings $\L$ and $\l (<2\pi)$,
which the standard model dynamics directly affect.
The remaining half is to select the vanishing (or tiny) value $\L_4=0$.

We note that
inclusion of another 3-brane at $r={\bar r}$ is straightforward.
The selection of $\L_4=0$ might be possible, for instance,
by a sector on a brane
\cite{Iza}
other than the brane at the origin.
For that purpose, we should also consider
the case for a negative bulk cosmological constant,
which will be investigated elsewhere.

\section*{Acknowledgments}

We would like to thank T.~Yanagida
for careful reading of the manuscript.

\end{document}